\documentclass[a4paper,11pt]{article}

\usepackage{amsmath,amssymb}
\usepackage{url,color}
\usepackage{graphicx}
\usepackage{multirow} % more control on table columns/rows
\usepackage{booktabs} % for better looking tables
\usepackage{parskip} % nice whiteline spacing between paragraphs
\usepackage{hyperref} % always load last

\makeatletter

\def\preprintno#1{\def\@preprintno{#1}}
\def\address#1{\def\@address{#1}}
\def\email#1#2{\thanks{\tt #1@{}#2}}
\def\abstract#1{\def\@abstract{#1}}
\renewcommand\abstractname{ABSTRACT}
\newlength\preprintnoskip
\newlength\abstractwidth
\@titlepagetrue

\renewcommand \maketitle{
	\begin{titlepage}
		\let \footnotesize \small
		\hfill\parbox{\preprintnoskip}{
		\begin{flushright}
			\@preprintno
		\end{flushright}
		}
		\hspace*{1cm} \vskip 60\p@
		\begin{center}
			{\Large\bf\boldmath \@title \par}\vskip 1cm
			{\sc\@author \par}\vskip 3mm
			{\@address \par}
		\end{center}
		\par \@thanks \vfill
		\begin{center}
			\parbox{\abstractwidth}{\centerline{\abstractname}%
			\vskip 3mm
			\@abstract}
		\end{center}
	\end{titlepage}
	\setcounter{footnote}{0}
	\let\thanks\relax\let\maketitle\relax
	\gdef\@thanks{}\gdef\@author{}\gdef\@address{}
	\gdef\@title{}\gdef\@abstract{}\gdef\@preprintno{}
}

\def\@citex[#1]#2{\if@filesw\immediate\write\@auxout{\string\citation{#2}}\fi
  \def\@citea{}\@cite{\@for\@citeb:=#2\do
    {\@citea\def\@citea{,\penalty\@m}\@ifundefined
       {b@\@citeb}{{\bf ?}\@warning
       {Citation `\@citeb' on page \thepage \space undefined}}%
\hbox{\csname b@\@citeb\endcsname}}}{#1}}
\def\citerange{\@ifnextchar [{\@tempswatrue\@citexr}{\@tempswafalse\@citexr[]}}
\def\@citexr[#1]#2{\if@filesw\immediate\write\@auxout{\string\citation{#2}}\fi
  \def\@citea{}\@cite{\@for\@citeb:=#2\do
    {\@citea\def\@citea{--\penalty\@m}\@ifundefined
       {b@\@citeb}{{\bf ?}\@warning
       {Citation `\@citeb' on page \thepage \space undefined}}%
\hbox{\csname b@\@citeb\endcsname}}}{#1}}

\long\def\@makecaption#1#2{%
  \vskip\abovecaptionskip
  \sbox\@tempboxa{#1: \emph{#2}}%
  \ifdim \wd\@tempboxa >\hsize
    #1: \emph{#2}\par
  \else
    \hbox to\hsize{\hfil\box\@tempboxa\hfil}%
  \fi
  \vskip\belowcaptionskip}
\makeatother

\newcommand{\subh}{\textit{\fontsize{1.5mm}{0mm}H}}

\newcommand{\limitHiggs}{607}
\newcommand{\limitEPWT}{405}
\newcommand{\limitHiggsEWPT}{694}
\newcommand{\limitDirectSearches}{638}

\begin{document}

%%%--- Title Page ---%%%
\preprintno{DESY 13-114}

\title{{\huge Little Higgs Model Limits from LHC} \\ \vspace{1mm} Input for Snowmass 2013}

\author{
	J\"urgen Reuter\email{juergen.reuter}{desy.de}$^a$, 
	Marco Tonini\email{marco.tonini}{desy.de}$^a$, 
	Maikel de Vries\email{maikel.devries}{desy.de}$^a$
}

\address{\it $^a$DESY Theory Group, D--22603 Hamburg, Germany}

\abstract{
	The status of several prominent model implementations of the Little Higgs paradigm, the Littlest Higgs with and without discrete T-parity as well as the Simplest Little Higgs are reviewed. For this, we are taking into account a fit of 21 electroweak precision observables from LEP, SLC, Tevatron together with the full $25 \, \textrm{fb}^{-1}$ of Higgs data reported by ATLAS and CMS. For the Littlest Higgs with T-parity an outlook on corresponding direct searches at the $8 \, \textrm{TeV}$ LHC is included. We compare their competitiveness with the EW and Higgs data in terms of their exclusion potential. This contribution to the Snowmass procedure contains preliminary results of \cite{RTV} and serves as a guideline for which regions in parameter space of Little Higgs models are still compatible for the upcoming LHC runs and future experiments at the energy frontier. For this purpose we propose two different benchmark scenarios for the Littlest Higgs with T-parity, one with heavy mirror quarks, one with light ones. \vspace{5mm}
}

\maketitle

%%%--- Introduction: the Little Higgs Paradigm ---%%%
\section{Introduction: the Little Higgs Paradigm}
\label{sec:intro}
The first run of LHC at $2 \, \textrm{TeV}$, $7 \, \textrm{TeV}$, and $8 \, \textrm{TeV}$ centre of mass energies has brought as main results the discovery of a particle compatible with the properties of the Standard Model (SM) Higgs boson as well as with electroweak precision tests (EWPT), and no significant excesses that could be traced to new particles or forces. All of these results have been highly discriminative in the sense of constraining the parameter space of models beyond the SM (BSM), whose focus has been to solve the problem of the quadratic sensitivity of the scalar Higgs mass term in the SM to any kind of UV physics. 

Scenarios that are weakly coupled at the TeV scale have already been favored over strongly coupled models by EWPT. There is however a class of models where new strong interactions are compatible with EWPT as a mechanism called collective symmetry breaking allows those models to stay weakly coupled at the TeV scale and moves their compositeness scale of strong dynamics up by an order of magnitude. The Higgs (and other scalar bosons) appear as pseudo Nambu-Goldstone bosons of a broken global symmetry very similar to QCD. The problem of quadratic divergences is solved as the Higgs itself is a composite state. Those models are known as Little Higgs models (for reviews see \cite{Schmaltz:2005ky,Perelstein:2005ka}, and references therein). There exists a huge variety of different implementations of the Little Higgs collective symmetry breaking mechanism, but all these models share several common features: an extended scalar sector compared to the SM Higgs, new heavy gauge bosons and new heavy fermions. In this Snowmass white paper we will focus on the three most common models, the Littlest Higgs \cite{ArkaniHamed:2002qy}, the Simplest Little Higgs \cite{Schmaltz:2004de} and the Littlest Higgs model with T-parity \cite{Cheng:2003ju}. 

The Littlest Higgs is the prime example of the product group models, while the Simplest Little Higgs is the corresponding front runner of the simple group models. In the product group models the EW gauge group is embedded into a product group structure, while the scalar sector is in an irreducible representation of the global structure group; for simple group models the EW group is embedded into a larger simple group, while the scalar sector is a reducible representation of the global symmetry structure. Both classes can in principle be discriminated by means of the EW quantum numbers of their additional scalar degrees of freedom \cite{Kilian:2004pp,Kilian:2006eh}. T-parity is a discrete symmetry that can (at least in principle) be introduced for any Little Higgs model. All new particles (with few exceptions) are odd under this parity, which avoids tree level contributions to the EW precision observables and improves the compatibility with EWPT by roughly an order of magnitude \cite{Csaki:2002qg,Kilian:2003xt,Hubisz:2005tx}. Furthermore, it offers the possibility of a dark matter candidate with the lightest T-odd particle (LTOP). However now, all new particles must be pair produced which also reduces the reach for direct searches. 

The most important parameter of Little Higgs models is the mass scale $f$ of the new particles, which coincides with the equivalent of the pion decay constant of the underlying nonlinear sigma model parameterizing the global symmetry breaking. For the sake of brevity we do not repeat the full formulae defining the three models under consideration, which can be found in \cite{RTV,Reuter:2012sd}, where also the full set of original references can be found. Other relevant model parameters will be specified in the corresponding sections. 

In this Snowmass white paper we will present the limits from the EWPT as well as the Higgs data from the two LHC collaborations ATLAS and CMS from the full 2011/2012 data sets for the Littlest Higgs model (L2H), the Simplest Little Higgs (SLH) and the Littlest Higgs model with T-parity (LHT). We also discuss search results for new particles recasted to the specifications of the LHT model (we concentrate on the LHT as the other models are constrained by Higgs data and EWPT to scales where direct searches at LHC8 are not feasible). Note that some of the direct search results are still partially preliminary and will be presented in full detail in \cite{RTV}. Section \ref{sec:ewpt} discusses the bounds from EWPT, section \ref{sec:higgs} the constraints from the LHC Higgs data. The recasted results from direct searches are discussed in section \ref{sec:direct} and in section \ref{sec:conclusion} a brief summary of the most important results is provided.

%%%--- Electroweak Precision Tests ---%%%
\section{Electroweak Precision Tests}
\label{sec:ewpt}
For the Electroweak Precision Tests (EWPT), we calculate the EW precision observables within the three different implementations of the Little Higgs models. In the L2H they arise from the tree level exchange of the heavy gauge bosons as well as from the vacuum expectation value (\emph{vev}) $v'$ of the Higgs triplet fields. Loops from heavy quarks to the EW gauge boson self energies are negligible. All of the observables can be expressed through the four variables $f$, $c \equiv \cos\theta$, $c' \equiv \cos\theta'$, and $x \equiv 4v' f / v^2$, i.e. the L2H scale, the mixing angles in the non-abelian and abelian gauge boson sector as well as the dimensionless ratio of the triplet and doublet \emph{vev}.

For the SLH model, the dominant contributions come from the $Z'$ boson corrections as well as the $Z-Z'$ mixing, and the parameter besides $f$ that goes in is the mixing angle between the two Goldstone boson multiplets, $t_\beta := \tan\beta$. Introducing T-parity removes all tree level contributions in the Littlest Higgs to the EW precision observables (except maybe the contribution of the T-even top quark to top quark operators). To the oblique parameters, only the T-even top contributes at 1-loop, as the T-odd top is an EW singlet. An additional contribution from the mirror fermions and a logarithmic correction to the oblique parameters from the modified Higgs couplings to the EW gauge bosons are included as well. There are negliglible pieces from the heavy gauge bosons to the EW precision observables. The corrections in the flavour sector in all three models can  actually be expressed by corrections to the EW charged and neutral current couplings from their SM values as an expansion in $v/f$. The explicit expressions of the corrections can be found in \cite{Reuter:2012sd}.  

\begin{figure}[!ht]
	\begin{center}
		\includegraphics[width=.48\textwidth]{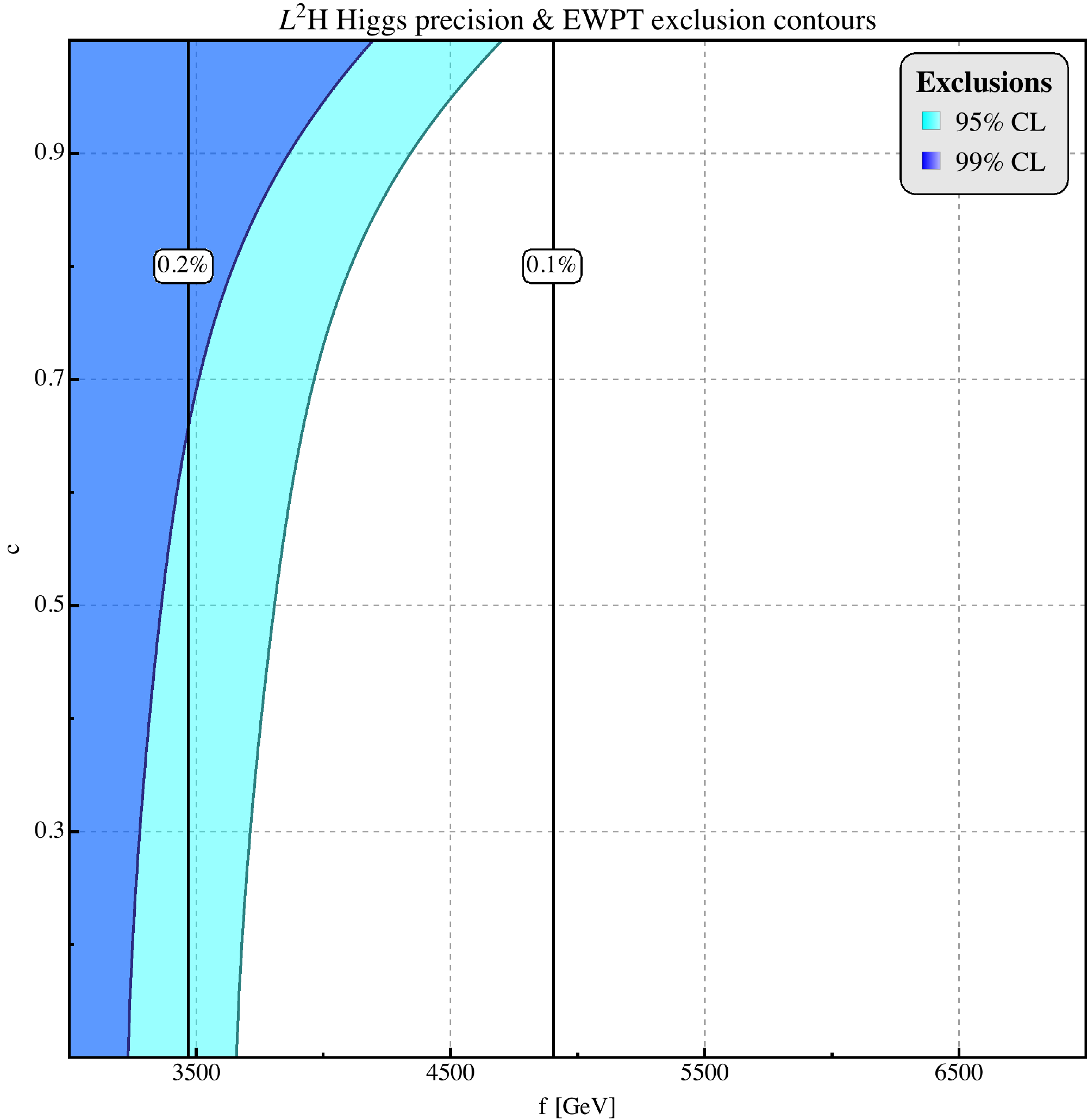}
    \hspace{.2cm}
		\includegraphics[width=.48\textwidth]{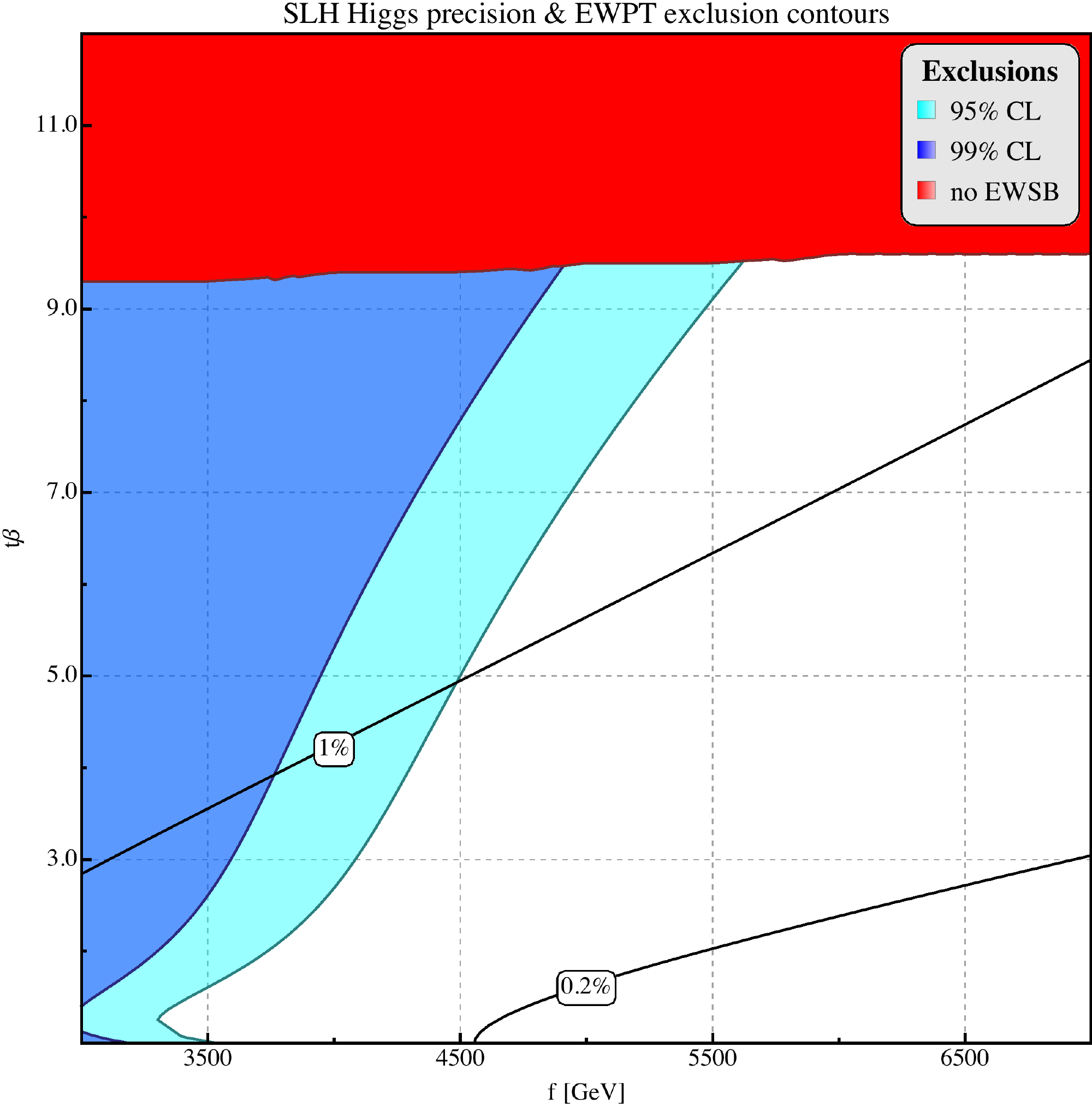} 
		\caption{Exclusion limits at $95\%$ CL (light blue) and $99\%$ CL (dark blue) from EWPT and Higgs data combined as a function of the Little Higgs model scale $f$. The thick black lines represent contours  of required fine-tuning. Left: Littlest Higgs model (L2H), showing the exclusion region for the mixing angle $c \equiv \cos\theta$ in the gauge sector; right: Simplest Little Higgs (SLH) exclusion region in $\tan\beta$, the mixing angle of the two nonlinear sigma model fields. The red region gives no viable EW symmetry breaking.} 
		\label{fig:Higgs_L2H_SLH}
	\end{center}
\end{figure}

For the calculation of the 21 EW precision parameters (low-energy data and on the $Z$ pole) we use the values given in \cite{Beringer:1900zz} taking the observed Higgs mass of 125 GeV into account. The values of the EW precision parameters are summarized in \cite{Reuter:2012sd}. For simplicity, we assume no correlations among the EW precision observables and include them in a $\chi^2$ fit defined just by quadratic deviations from the measured values divided by the (squared) experimental uncertainties. 

From the EWPT, one gets the following exclusion limits at $95\%$ CL on the Little Higgs scale for the corresponding models: $f \lesssim  5100 \, \textrm{GeV}$  for the L2H, $f\lesssim 3700 \, \textrm{GeV}$ for the SLH, and $f \lesssim \limitEPWT \, \textrm{GeV}$ for the LHT. Exclusion plots can be found in \cite{RTV}, here they are only shown in combination with the Higgs data in the next section.

%%%--- Constraints from Higgs data ---%%%
\section{Constraints from Higgs data}
\label{sec:higgs}
After the LHC discovery of the Higgs like boson compatible with EWPT, the precision of the data (besides systematic uncertainties) grows with the square root of the integrated luminosity and is starting being competitive with the discriminative power of the EWPT itself \cite{ATLAS:2013sla,CMS:yva}. The two collaborations, ATLAS and CMS, express their results for the Higgs channels in terms of the signal strength modifier $\mu$, which for a given Higgs mass, $m_{h}$, is the ratio between the observed cross section times branching ratio folded with the experimental efficiency normalized to the corresponding values of the SM. 

For the $\chi^2$ analysis of the compatibility of the Higgs data with Little Higgs models we take all published experimental data into account, coming from the vector boson channels $\gamma\gamma$ (all different experimental categories), $ZZ$, $WW$, as well as the two fermionic channels $bb$ and $\tau\tau$. We do not repeat the collection of all the values here, the details can be found in \cite{RTV,Reuter:2012sd}. The $95\%$ and $99\%$ CL regions are then defined by the cumulative distribution function for an appropriate number of degrees of freedom (DOF) within our $\chi^2$ measure. For all channels we take the latest public $7$ and $8 \, \textrm{TeV}$ data of the 2011 and 2012 LHC runs. 

\begin{figure}[!ht]
	\begin{center}
		\includegraphics[width=.48\textwidth]{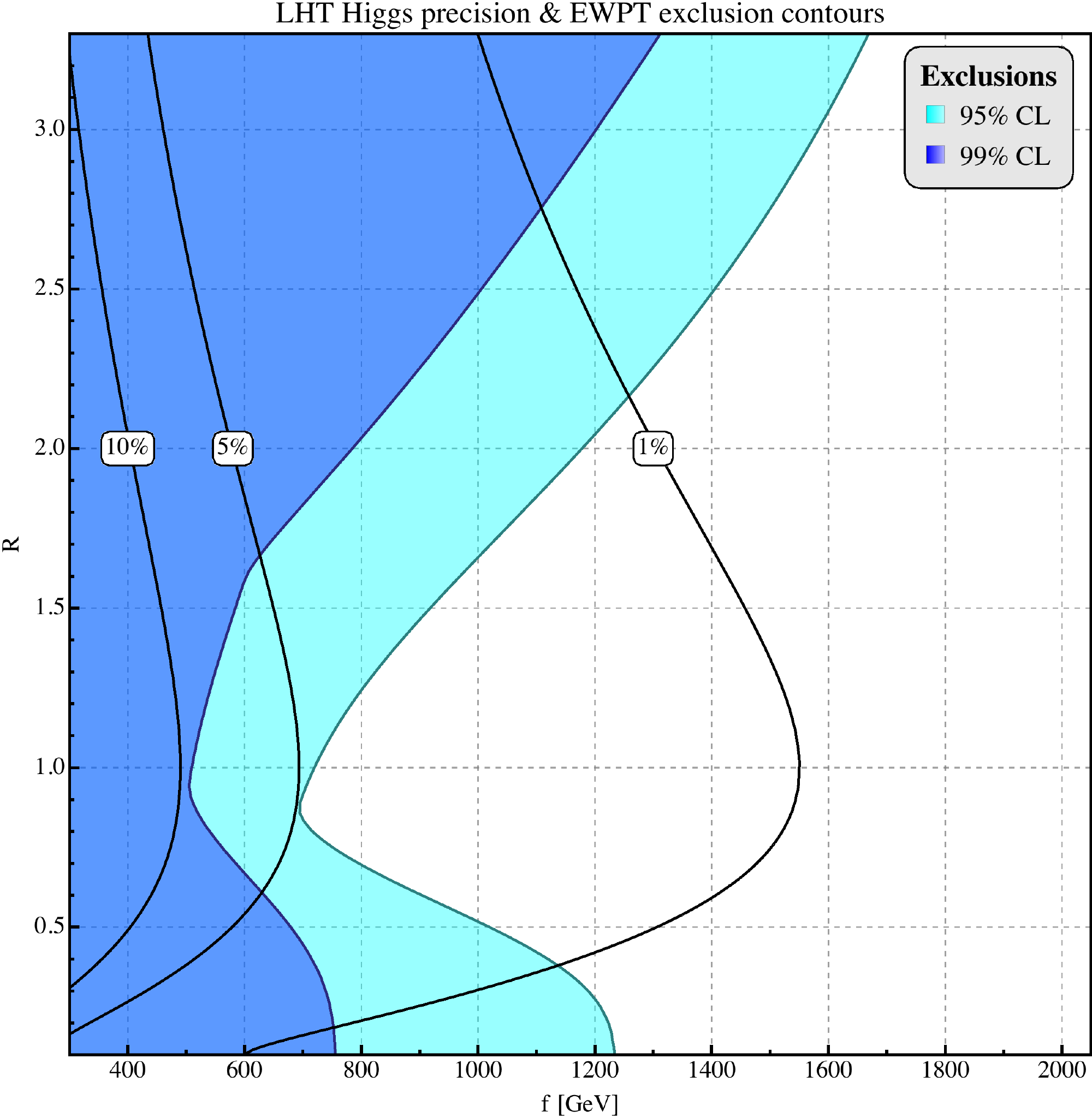} 
		\caption{Exclusion regions for the LHT model from EWPT and Higgs data combined, for the ratio $R = \lambda_1/\lambda_2$ of two Yukawa couplings in the T-parity doubled fermion sector. Colour code as in figure \ref{fig:Higgs_L2H_SLH}.} 
		\label{fig:Higgs_LHT}
	\end{center}
\end{figure} 

Our results are shown in figure \ref{fig:Higgs_L2H_SLH} for the L2H (left) and the SLH model (right) in the $(f,c)$ and $(f,t_\beta)$ plane, respectively as well as for the LHT in figure \ref{fig:Higgs_LHT} in the $(f,R)$ plane. The black lines are contour lines of the fine tuning variable $1/\Delta$, with $\Delta=|\delta \mu^2| / \mu_{\text{obs}}^2$, $\mu_{\text{obs}}^2= \frac{m_h^2}{2}$. Here, $\mu$ is the Higgs mass parameter and $\delta\mu$ its leading radiative correction. This means, a worse fine-tuning results in a smaller value of $1/\Delta$, as a certain parameter value has to be tuned to a much smaller level to achieve the experimentally observed Higgs mass. The $95\%$ and $99\%$ CL exclusion regions combining the Higgs and EW are shown in light blue and dark blue, respectively. Due to the strong contributions to the EW precision observables, in the case of the L2H and the SLH the EWPT dominate the exclusion region, and the inclusion of Higgs data compatible with the SM (in the large-$f$ region) just reduces the $\chi^2$/DOF. This leads to a weaker exclusion from the Higgs plus EWPT combination as from the EWPT data alone: the $95\%$ CL exclusion for the L2H is $f \lesssim 3700 \, \textrm{GeV}$, while it is $f\lesssim 3300 \, \textrm{GeV}$ for the SLH. For the LHT, the general limit (independent of $R$ \cite{Reuter:2012sd}) improves to a $95\%$ CL exclusion of $f \lesssim \limitHiggsEWPT \, \textrm{GeV}$, while the constraint from Higgs data alone would be $f \lesssim \limitHiggs \, \textrm{GeV}$ at $95\%$ CL. This comes mainly from the fact that the Higgs exclusion for the LHT is independent from $R$ and does not suffer from the dip around $R \approx 1$ coming from a cancellation of LHT contributions to the $T$ parameter.

%%%--- Limits from Direct Searches ---%%%
\section{Limits from Direct Searches}
\label{sec:direct}
In this section, we take into account different searches published by the LHC experiments that have been optimized especially for low-energy supersymmetry. As the bounds from EWPT (and also Higgs data) for the L2H and SLH already push up the symmetry breaking scale $f$ into the multi-TeV region, this makes direct discoveries of new particles at $8 \, \textrm{TeV}$ impossible and even at $14 \, \textrm{TeV}$ very difficult. On the other hand, most of the direct searches for the L2H and SLH could be covered with the $Z'$, $W'$ searches as well as the searches for heavy vector-like quarks. For these reasons, we focus on the LHT for the direct searches. 

\begin{figure}[!ht]
	\begin{center}
		\includegraphics[scale=0.55]{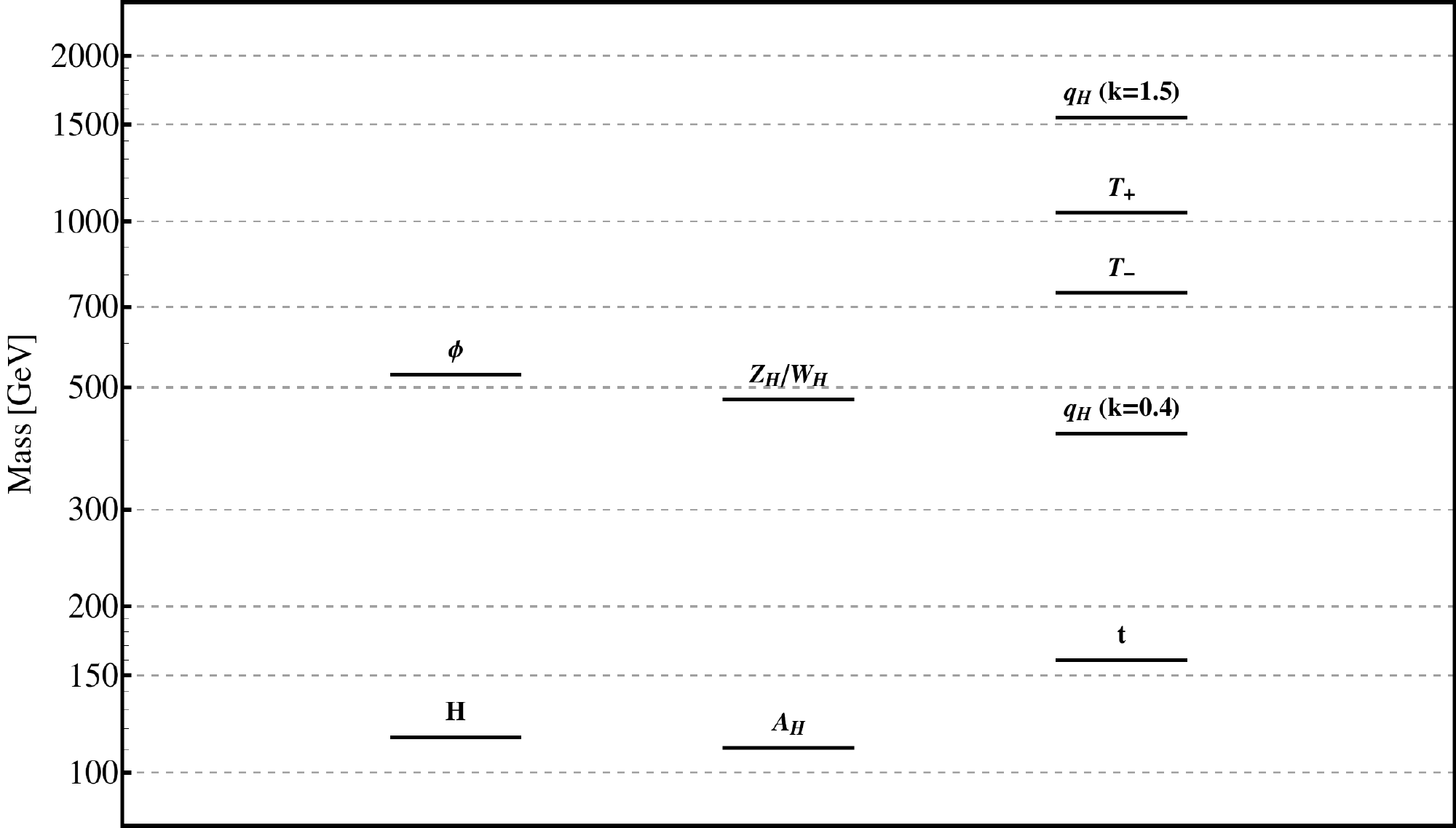}
		\caption{LHT benchmark scenarios, fixing $R=1.0$ in order to minimize EWPT: heavy mirror quarks ($k=1.5$) or rather light ones ($k=0.4$).} 
		\label{fig:massspectrum}
	\end{center}
\end{figure}

The LHT model can be parameterized in principle by two parameters beyond those of the SM and the Little Higgs scale $f$: the ratio of Yukawa couplings in the top sector, $R := \lambda_1/\lambda_2$, as well as a parameter from the sector of mirror fermions necessary to implement T-parity: a Yukawa coupling $k$ that sets the scale of the mirror fermions. Generically, the heavy photon $A_H$ is the lightest T-odd particle (LTOP), while the heavy vectors $W_{\subh}$ and $Z_{\subh}$ are degenerate up to corrections of order $v^{2}/f^{2}$ as demanded by EWPT. For our studies, the scalar triplet can be ignored. The top partners $T_{+}, T_{-}$ are always heavier than all bosons. The parameter $k$ determines the mass of the quark partners: for $k \gtrsim 0.45$ the quark partners are heavier than all gauge bosons, opening up  the decay $q_{\subh} \rightarrow V_{\subh} \; q$. For lower values of $k$ there is only the decay channel $q_{\subh} \rightarrow A_{\subh} \; q$. In the region $0.1 \lesssim k \lesssim 0.2$ the $q_{\subh}$-$A_{\subh}$ mass difference is rather small, making the spectrum compressed. In figure \ref{fig:massspectrum} we propose two benchmark scenarios for the LHT, fixing $R \equiv 1.0$, thereby minimizing the contributions to the EW precision observables. The two different cases of decay patterns for the quark partners are reflected in the two different choices for $k$, $k=1.5$ or $k=0.4$. Tables with the dominant decay modes for the different states in the two benchmark cases can be found in reference \cite{RTV}.

In order to discuss direct search limits, we briefly discuss the major production modes at the LHC. The T-even top partner $T_{+}$ is the only new state to be singly produced, while all other particles have to be pair produced, highly reducing the available phase space with increasing masses. Due to the PDF enhancement, the pair production of the quark partners $q_{\subh}$ could be significant, especially if their masses are not too large. Their production is mainly dominated by QCD processes, but also EW processes with a heavy $W_{\subh}$ in the $t$-channel significantly contribute. All details about the cross sections and listings of the relevant signatures can be found in \cite{RTV}. 

\begin{figure}[ht]
	\begin{center}
		\includegraphics[scale=0.6]{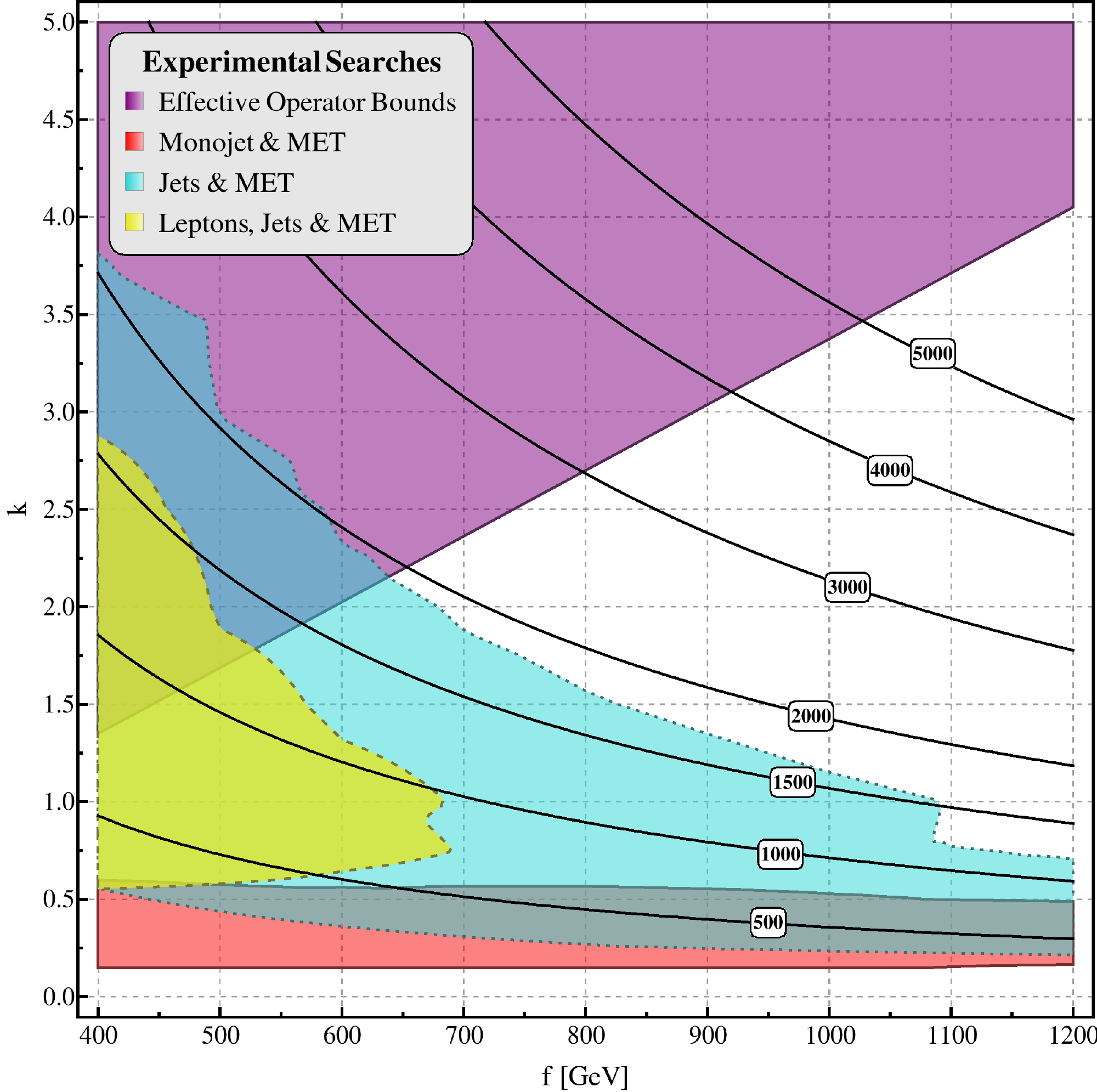} \hspace{.2cm}
		\caption{Exclusion limits from direct searches at LHC8 displayed in the $(f,k)$ plane. The different categories comprise limits from operator bounds and searches from monojets, jets and leptons plus jets. The contour lines show the mass of the heavy quark partners.}  
		\label{fig:exclusionlimitsfk}
	\end{center}
\end{figure}

Before we discuss direct searches at the LHC, there are constraints coming from integrating out the T-odd quark generating effective four-fermion operators, $\mathcal{O}_{\text{4-f}} = - \tfrac{k^2}{128 \, \pi^2 f^2} \bar{\psi}_L \gamma^{\mu} \psi_L \bar{\psi}^{\prime}_L \gamma_{\mu} \psi^{\prime}_L$ + $\mathcal{O} \left( \frac{g}{k} \right)$, under the assumption that $k$ is diagonal and flavour independent. $\psi$ and $\psi^{\prime}$ are different SM fermions. Experimental bounds on four-fermion interactions provide an upper bound on the \emph{T}-odd fermion masses, translating into an upper bound on $k$. At LHC, angular distributions and rates in dijet events lead to bounds of the order $\Lambda = 15 \, \textrm{TeV}$ for (for constructive interference with the SM). The most stringent bounds, however, are still from LEP, constraining the coefficient of the operator $eedd$ operator \cite{Hubisz:2005tx,Beringer:1900zz}. This leads to the upper bound $k^2 \lesssim 0.367 \, \pi^3 \, \left( f / \textrm{TeV} \right)^2$. This bound is shown in the total exclusion plot in figure \ref{fig:exclusionlimitsfk} as the triangular shaped purple region in the upper half. We do expect LHC to improve these bounds, since $8 \, \textrm{TeV}$ analyses for operator bounds of this form still need to be published. Moreover many promising results from the $14 \, \textrm{TeV}$ run can be expected in this area, probably driving limits up to $\mathcal{O}(30 \, \textrm{TeV})$.  

Next, we will consider all searches from the $8 \, \textrm{TeV}$ LHC run relevant to the Littlest Higgs with T-parity. For each analysis, we identify the final states and the contributing production modes in the LHT model. Typically, the production cross section times branching ratio ($\sigma \times \mathrm{Br})$ depends on the scale $f$ and either the parameter $k$ or $R$. Therefore exclusion limits can either be presented as contours in the $(f,k)$ or the $(f,R)$ plane. For this Snowmass white paper, we summarize the exclusion limits in $(f,k)$ plane in figure \ref{fig:exclusionlimitsfk}. This allows to set a lower limit on the scale $f$. In the following, we briefly discuss the different searches. More details can be found in \cite{RTV}. 

To obtain the exclusion limits from recasting the LHC analyses (mostly for SUSY searches) we generated signal events for the LHT model with MadGraph \cite{Alwall:2011uj} and WHIZARD \cite{Kilian:2007gr,Moretti:2001zz}. The events are then processed with the fast detector simulation Delphes 3.0 \cite{deFavereau:2013fsa} to simulate either the ATLAS or CMS detector in order to evaluate the cut efficiencies of the different analyses applied to the assumed LHT signal. With these events, we have recasted three different classes of searches, monojets and missing transverse energy (MET), jets and MET, and jets, leptons and MET. 

\paragraph{Monojet \& MET:}
Both ATLAS and CMS have presented monojet searches with $8 \, \textrm{TeV}$ data for final states containing a single energetic jet and missing transverse energy \cite{ATLAS:2012zim,CMS:rwa}. They are quite similar, requiring a single energetic jet and at most two further jets with $p_T > 30 \, \textrm{GeV}$. Furthermore, a lepton veto is imposed and the presence of significant missing transverse energy is required in both analyses. For more details on the signal regions, see \cite{RTV}. Additional suppression of QCD dijet background is done differently: ATLAS requires the azimuthal separation between the direction of missing $E_T$ and the second leading jet (if present) to be greater than $0.5$, whereas CMS only keeps two jet events if the azimuthal separation between the jets is less than $2.5$.

In the absence of any deviation from the SM, both experiments quote $95\%$ CL upper bounds on the signal cross section times efficiency for their corresponding signal regions. Both monojet searches are suitable for final state topologies containing one or two hard jets and missing transverse energy. Hence both LHT production modes $p \, p \to q_\subh \, q_\subh$ and $p \, p \to q_\subh \, A_\subh$ may contribute, provided the heavy quark partner decays to a quark and a heavy photon $q_\subh \to A_\subh \, q$. Therefore these searches are mostly sensitive in the parameter regions where quark partners $q_\subh$ are lighter than the gauge bosons $W_\subh$ and $Z_\subh$, i.e. for lower values of $k$ and independent of $f$, such that they entirely decay as $q_\subh \to A_\subh \, q$. Figure \ref{fig:exclusionlimitsfk} shows the excluded regions recasting both ATLAS and CMS monojet plus MET analysis as a red band at the lower end.

\paragraph{Jets \& MET:}
In this second category, all searches with at least two jets, missing transverse energy and no leptons in the final state are investigated. Numerous searches interpreted in terms of SUSY have been presented for the $8 \, \textrm{TeV}$ data by ATLAS \cite{ATLAS:2012ona,ATLAS:2013cma} and CMS \cite{Chatrchyan:2013lya}. The first ATLAS search is optimized for squarks and gluinos, the second one for stops, whereas the CMS search looks more generically at squarks, sbottoms and gluinos. In the LHT scenario these searches can be applied to pair production of heavy gauge bosons and heavy quarks or associated productions like $V_\subh \, q_\subh$. In those cases, the heavy quarks decay to heavy vectors (decaying to SM gauge bosons, all hadronic final states, and MET from the $A_\subh$) and quarks. Also covered are possible LHT production modes like pair production of the T-odd top $t_\subh$ with subsequent stop-like decay, or two heavy quarks where at least one decays like $q_\subh \to Z_\subh \, q$, giving the required two b-jets via $Z_\subh \to H \, A_\subh \to bbA_\subh$. 

Again, each of the above searches provide $95\%$ CL limits on the cross sections times efficiencies in the absence of any signal. Also, the CMS analysis \cite{Chatrchyan:2013lya} can be recast as a search for pair production of heavy gauge bosons and quarks as well as the associated productions $V_\subh \, q_\subh$ with completely hadronic final states. A detailed list of the experimental signal regions, the SM background suppression, the cut flows and the efficiencies and the recast procedure can be found in \cite{RTV}. The results from recasting the ATLAS and CMS searches from jets plus MET described above for the LHT model are shown as light blue exclusion region in figure \ref{fig:exclusionlimitsfk}. 

\paragraph{Leptons, Jets \& MET:}
Here, all searches involving tagged leptons, at least two jets and missing transverse energy are collected. Some of the jets may be b-tagged. For this purpose, we consider the following searches by ATLAS and CMS resulting in corresponding constraints for the LHT model: \cite{ATLAS:2012tna,ATLAS:2013pla,ATLAS:2013tma}. These searches match also LHT topologies like production of two heavy quarks, which then decay to heavy gauge bosons $q_\subh \to W_\subh \, q$ or $\to Z_\subh \, q$ with one or both $W$s decaying leptonically. $b$-jets may arise from the Higgs decay in the $Z_\subh \to H \, A_\subh$ decay chain, or from the T-odd top partner decay chain $t_\subh \to t \, A_\subh$. Another signature covered is pair production of same charge heavy quarks $p p \to q_\subh q_\subh$ with subsequent decays into gauge boson partners $W_\subh$ with now leptonic decays for the $W$s. 

From the searches with leptons, jets and MET described in this paragraph, $95\%$ CL exclusion limits in the $(f,k)$ plane can be extracted. The results from the recast are presented in figure \ref{fig:exclusionlimitsfk} as yellow region in the middle of the left part of the plot. From our recast, we conclude that searches for both a single and two leptons perform equally, as long as no $b$-jets in the final states are required. 

\paragraph{Combination:}
Finally, we combine all direct searches in figure \ref{fig:exclusionlimitsfk}, derive a combined exclusion on the scale $f$ from it, and compare with the bounds from EWPT and Higgs data. We get an exclusion limit at $95\%$ CL from direct searches of $f \lesssim \limitDirectSearches \, \textrm{GeV}$, which is still less than the combination of EWPT and Higgs data.

%%%--- Conclusion ---%%%
\section{Conclusion}
\label{sec:conclusion}
We summarized constraints from EW precision test (EWPT), the Higgs data from the ATLAS and CMS analyses for the three most economic implementations of Little Higgs models, the Littlest Higgs (L2H), the Simplest Little Higgs (SLH) and the Littlest Higgs with T-parity. For the latter, we also recasted direct searches from the LHC experiments optimized for SUSY searches. While the constraints from EWPT as well as from the Higgs data for L2H and SLH are in the regime above $3 \, \textrm{TeV}$, LHT is in much better shape. Limits on the symmetry breaking scale $f$ in LHT are
\begin{equation*}
	\boxed{f \gtrsim \limitHiggsEWPT \, \textrm{GeV}} \quad \textrm{at} \quad 95\% \textrm{ CL} ,
\end{equation*}
from the combination of EPWT and Higgs data and
\begin{equation*}
	\boxed{f \gtrsim \limitDirectSearches \, \textrm{GeV}} \quad \textrm{at} \quad 95\% \textrm{ CL} ,
\end{equation*}
from direct searches. These exclusion limits arise from the combination of monojet \& MET, jets \& MET, leptons, jets \& MET and bounds on contact interactions. They are in the same ballpark as EWPT plus Higgs data but are not completely competitive yet. 

As we have not yet completed our studies for the deadline of the Snowmass process, this study is just a proceedings style summary. Specifically, we do not give any strategies for the optimization of the cut flows for the direct search analysis which will be found in \cite{RTV}. This would enable the LHC experiments to improve on their limits for the Little Higgs parameter spaces and would also allow to give a more reliable estimate for the prospects for $14 \, \textrm{TeV}$, the high luminosity LHC upgrade (HL-LHC), or possible future energy upgrades to $33$ or $100 \, \textrm{TeV}$. Such estimates do however rely on the possible calibration of the experiments. Furthermore, prospects for a future ILC are also not included in this note (however, see \cite{Baer:2013cma}).

%%%--- Acknowledgements ---%%%
\section*{Acknowledgements}
The authors like to thank Diptimoy Ghosh, Paolo Gunnellini, Maxim Perelstein, Kazuki Sakurai and Andreas Weiler for valuable remarks and discussions. MT and MdV have been partially supported by the Deutsche Forschungsgemeinschaft within the Collaborative Research Center SFB 676 "Particles, Strings, Early Universe".

%%%--- Bibliography ---%%%
\bibliographystyle{unsrt}

%%%--- End document ---%%%
\end{document}